# NMR INVESTIGATION OF THE ORGANIC CONDUCTOR λ-(BETS)$_2$FeCl$_4$

W.G. Clark[1], Guoqing Wu[1], P. Ranin[1], L.K. Montgomery[2], and L. Balicas[3]

[1]University of California at Los Angeles (UCLA), Department of Physics and Astronomy
[2]Indiana University, Bloomington, Indiana, Department of Chemistry
[3]National High Magnetic Field Laboratory (NHMFL), Tallahassee, FL

Short title: NMR INVESTIGATION OF λ-(BETS)$_2$FeCl$_4$

Corresponding author:

Professor W. Gilbert Clark
UCLA Department of Physics and Astronomy
Box 951547, Los Angeles, CA 90095-1547, USA
Tel: (310) 825-4079; Fax: 825-5734
email: clark@physics.ucla.edu (or wgclark@ucla.edu)

ABSTRACT

The two-dimensional organic conductor λ-(BETS)$_2$FeCl$_4$ has an unusual phase diagram as a function of temperature and magnetic field that includes a paramagnetic metal (PM) phase, an antiferromagnetic insulating (AFI) phase, and a field-induced superconducting phase [S. Uji, H. Kobayashi, L. Balicas, and James S. Brooks, Adv. Mater. **14**, 243 (2002), and cited references]. Here, we report a preliminary investigation of the PM and AFI phases at 9.0 T over the temperature range 2.0-180 K that uses proton NMR measurements of the spectrum, the spin-lattice relaxation rate ($1/T_1$), and the spin echo decay rate ($1/T_2$). The sample is a small single crystal whose mass is ~3 μg (approximately $2 \times 10^{16}$ protons). Its small size creates several challenges that include detecting small signals and excluding parasitic proton signals that are not from the sample [H.N. Bachman and I.F. Silvera, J. Mag. Res. **162**, 417 (2003)]. These strategies and other techniques used to obtain viable signals are described.

1. INTRODUCTION

The two-dimensional organic conductor λ-(BETS)$_2$FeCl$_4$ has an unusual phase diagram as a function of temperature and magnetic field that includes a paramagnetic metal (PM) phase, a (canted) antiferromagnetic insulating (AFI) phase, and a field-induced superconducting phase [1]. Here, we report a preliminary investigation of the PM and AFI phases at 9.0 T over the temperature range 2.0-180 K that uses proton NMR measurements of the spectrum, the spin-lattice relaxation rate ($1/T_1$), and the spin echo decay rate ($1/T_2$).

Because the sample is a small, fragile single crystal whose mass is only ~3 μg (approximately $2 \times 10^{16}$ protons), there are several experimental challenges to overcome. They include using a very small NMR coil and excluding parasitic proton signals that are not from the sample [2]. These strategies and other techniques used to obtain viable signals are described in detail in Section 2. S. Endo et al. [3] have reported the NMR spectrum for an aligned 3 mg cluster of needles in a magnetic field $\boldsymbol{B}_0 \approx 2$ T aligned perpendicular to the $c$-axis.

Three main experimental results are presented in Section 3 along with a brief discussion of their main features. The first is that as the temperature is lowered from 180 K to 4.2 K, the proton NMR spectrum obtained with the field aligned parallel to the $a$-$c$ plane shows an increase in width from approximately 0.9 MHz to 9 MHz and the development of numerous structural features. This behavior shows that the proton NMR is a useful probe of the local magnetic field structure of the $Fe^{+3}$ moments. Below about 4 K, which corresponds to the transition to the AFI phase, the width of the spectrum stays the same, but its structural features become significantly less distinct. The reason for this behavior has not yet been analyzed in detail.

The second experimental result is that $1/T_2$, which is approximately flat over the range 4-70 K, has a rapid drop below about 4 K. This behavior is attributed to a change in the low frequency dynamics of the magnetic fluctuations that occurs upon going from the PM to the AFI phase. Finally, $1/T_1$ has a temperature ($T$) dependence below about 100 K that indicates an energy-activated mechanism which is probably associated with the saturation of the $Fe^{+3}$ moments at low $T$ and high $B_0$. Since this work is still in progress, a more complete discussion will be published later.

2. SAMPLES AND EXPERIMENTAL METHODS

Single crystal λ-(BETS)$_2$FeCl$_4$ samples were prepared as described by Montgomery et al. [4]. The dimensions of the sample used for the measurements reported here were $(1.2 \pm 0.1)$ mm $\times (0.065 \pm 0.010)$ mm $\times (0.018 \pm 0.005)$ mm, which corresponds to $(3.8 \pm 1.8)$ μg in mass and $(2.7 \pm 1.3) \times 10^{16}$ protons.

Because of the small size of the samples, very small coils had to be constructed to obtain a filling factor that provided a viable sensitivity, which also resulted in obtaining a large rf magnetic field with relatively low power [5, 6]. The NMR coil was 40 turns of 25 μm diameter bare copper wire wound on a 75 μm diameter wire form. The individual turns were separated by about 10 μm. After being wound, the coil was slipped off the form and soldered to a twisted pair of 125 μm diameter Teflon-insulated Cu wires that attached the coil to the rest of the NMR coil circuit. The sample was placed in the coil and held in place with a very small amount of Apiezon grease at each end. It was oriented by eye to align $\boldsymbol{B}_0$ within approximately ± 5 deg of being perpendicular to the $c$-direction and parallel to the $ac$-plane.

Several steps were taken to reduce the spurious proton signals that are added to the total observed signal when the sample has a small signal. Insofar as possible, all materials rich in protons were kept a long distance from the coil, its leads, and other parts of the circuit whose rf currents were capable of producing a significant rf magnetic field. In addition, the amount of materials nominally free of protons, such as Teflon [2], were avoided or used in the smallest amount possible. Before inserting the sample, the coil and its leads were cleaned in an ultrasonic acetone bath to remove grease and other soluble products containing protons from the coil. It was also helpful to identify spurious signals by placing the sample on the edge of the homogeneous part of the magnetic field so that spurious signals spatially separated from the sample were shifted in their NMR frequency. Another important step that helped to suppress spurious proton signals was to do all the measurements with a spin echo, which discriminated against spurious signals generated in regions with a smaller value of the rf magnetic field ($B_1$) than that inside the NMR coil. This step was especially useful because some of the spurious signals had a relatively large signal from the free induction decay but a very small one from the spin-echo. Also, the value of $1/T_1$ for the protons in the sample was very short (see Section 3), which helped to distinguish them from those responsible for the spurious signals, which tended to have a much smaller value of $1/T_1$. As shown in Section 3, when these steps were taken the intensity of the spurious proton signals was reduced to less than 4 % of that from the λ-(BETS)$_2$FeCl$_4$ sample.

A laboratory-built helium gas flow system was used to provide the temperatures (2-180 K) used for the measurements. The NMR spectrometer was a laboratory-built model operating in the range near 383 MHz. Because the NMR spectrum had a width of up to 12 MHz (2.8 kG), short rf pulses and a wide receiver bandwidth (± 1 MHz) were used to record the spin-echo signals. For most of the measurements, the pulse sequence that optimized the height of the spin echo used to record the NMR signal was a 0.2 μs π/2 pulse ($B_1$ = 294 G, 1.25 MHz proton frequency) followed by a 0.3 μs pulse separated by a time interval τ. Because of the small volume of the NMR coil, a transmitter power of only 20 W was required. Phase cycling was used to remove ringing and other artifacts from the echo signal. To obtain a viable signal-to-noise ratio each echo signal was averaged 2,000 times at 180 K and 128 times at 4.2 K and lower temperatures.

Proton NMR spectra were obtained using the frequency stepped and summed (FSS) method [7]. Because of the wide bandwidth of the receiver and the short pulses used to obtain the spectrum, the typical frequency step for each acquisition was 0.2-0.5 MHz. A Fourier transform of the full echo was used to obtain the spectrum at each frequency. The question of how large a frequency range could be covered without retuning the probe was determined by monitoring the echo height while varying the resonant frequency of the probe. By using the criterion that the echo height should be reduced by less than 15 % because of probe detuning, it was found that a frequency range of 4 MHz could be used without retuning the probe. Since some of the frequency sweeps used covered 24 MHz, the probe had to be retuned only 6 times during such measurements.

Measurements of $1/T_1$ were made using a 0.2 μs π/2 pulse to prepare the initial state with a very small magnetization followed by spin-echo read sequence after the recovery time $t$. The integral of the echo, which corresponds to the Fourier transform of the signal at the measurement frequency [7] was used for the amplitude of the signal. Since the form of the recovery curve indicated a fairly broad distribution of relaxation rates, it was fitted to a stretched exponential function. In this case, the value obtained for $T_1$ corresponds to the time for the recovery curve to reach $1/e$ of its final value.

Measurements of $1/T_2$ were made by applying the 0.2 -τ- 0.3 μs spin-echo pulse to the equilibrium magnetization and recording the integral of the echo as a function of the pulse separation time τ. The value used for $T_2$ was the time for the echo integral to decay to $1/e$ of its value extrapolated to τ = 0.

3. EXPERIMENTAL RESULTS

The three experimental results presented in this Section and their qualitative interpretations are the NMR Spectrum, $1/T_1$, and $1/T_2$.

NMR SPECTRUM - The NMR spectra obtained at several temperatures between 2 K and 180 K are shown in Fig. 1, where the normalized proton absorption spectrum is plotted as a function of the shift from 383.0 MHz for several values of $T$. The magnetic field value is 9.000 T aligned perpendicular to the $c$-axis in the $ac$-plane. There, it is seen that as the temperature is lowered from 180 K to 4.2 K in the PM phase, the spectrum broadens and develops six main peaks that show an additional weak structure at lower $T$. As the sample is cooled further into the AFI phase, the details of the spectrum become somewhat smeared and an additional peak at the high frequency side of the spectrum appears.

A measure of the width of the spectrum is indicated by the arrows on Fig. 1. They correspond to the frequencies at which the intensity is 20 % of its maximum value. The width defined in this way is fits rather well the susceptibility of this material, which follows the Curie Law with a Curie temperature of 10 K [8] (not shown here). Because of the large proton shifts and their temperature dependence, we attribute the behavior of the spectra to the local field of the $Fe^{3+}$ electron spin moments at the different proton sites in the λ-(BETS)$_2$FeCl$_4$ structure.

Figure 2 shows the sample (upper curve) and the parasitic proton spectra (lower curve) at 20 K using the same gain for both signals. The signal for the spurious proton signal is obtained by removing the sample and repeating the spectrum measurement and the signal from the sample is obtained by subtracting the measured spurious signal from the total signal. Because of this subtraction, the noise for the sample signal is about 40 % larger than for the spurious one. When analyzed in this way, the number of spins in the spurious signal obtained from the integral of the absorption spectrum is less than 4 % of that of the sample. Thus, in these measurements the parasitic proton signal has been reduced to an insignificant size.

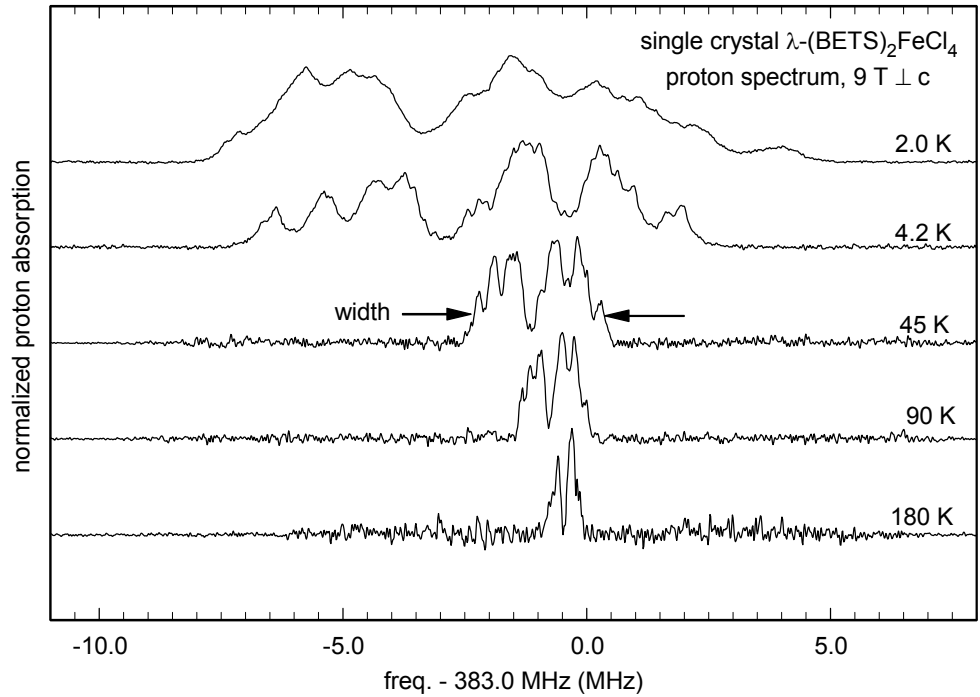

Fig. 1: Shift of the proton NMR spectrum from 383.0 MHz for $B_0 = 9.000$ T in the *a-c* plane perpendicular to the *c*-axis at several temperatures from 180 K to 2 K for a single crystal of $\lambda$-(BETS)$_2$FeCl$_4$.

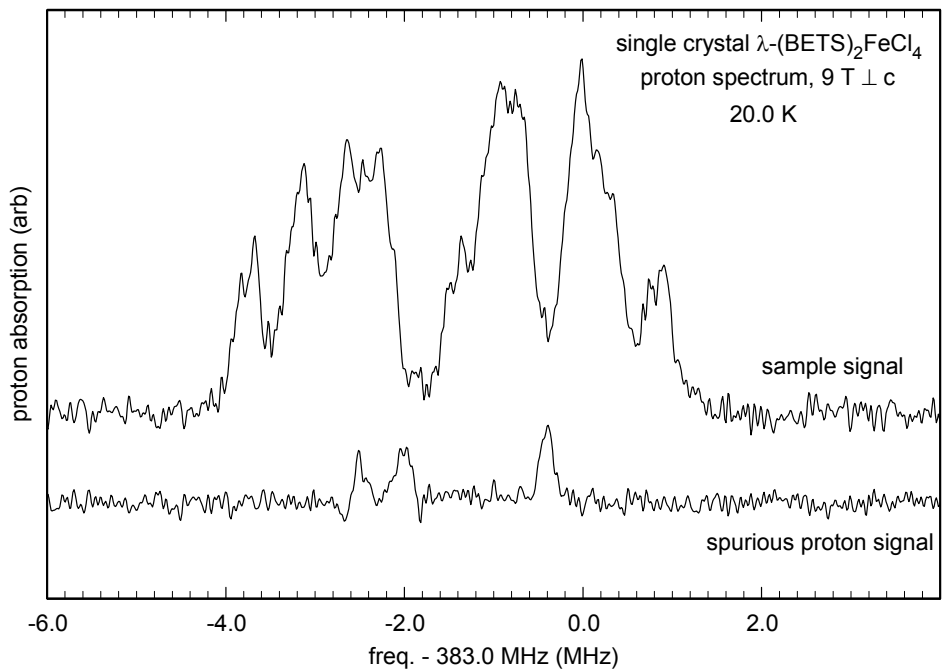

Fig. 2: Proton absorption signal from the sample only (upper curve) and the spurious proton absorption signal with the sample removed from the coil (lower curve) at 20 K for $B_0 = 9.000$ T in the *a-c* plane perpendicular to the *c*-axis.

SPIN-LATTICE RELAXATION - The stretched-exponential analysis of the proton spin-lattice relaxation rate $1/T_1$ as a function of $T$ is shown in Fig. 3. The error bars indicate the uncertainty in the analyzed results, which becomes fairly large at high $T$ because of the relatively low signal-to-noise ratio. Since there is a large width and shift of the line, the measurement frequency was set to the center of the first large peak to the right of the minimum in the spectrum (see Fig. 1). The solid line shows a temperature variation that corresponds to an energy gap behavior with an energy gap $\Delta = 9$ K.

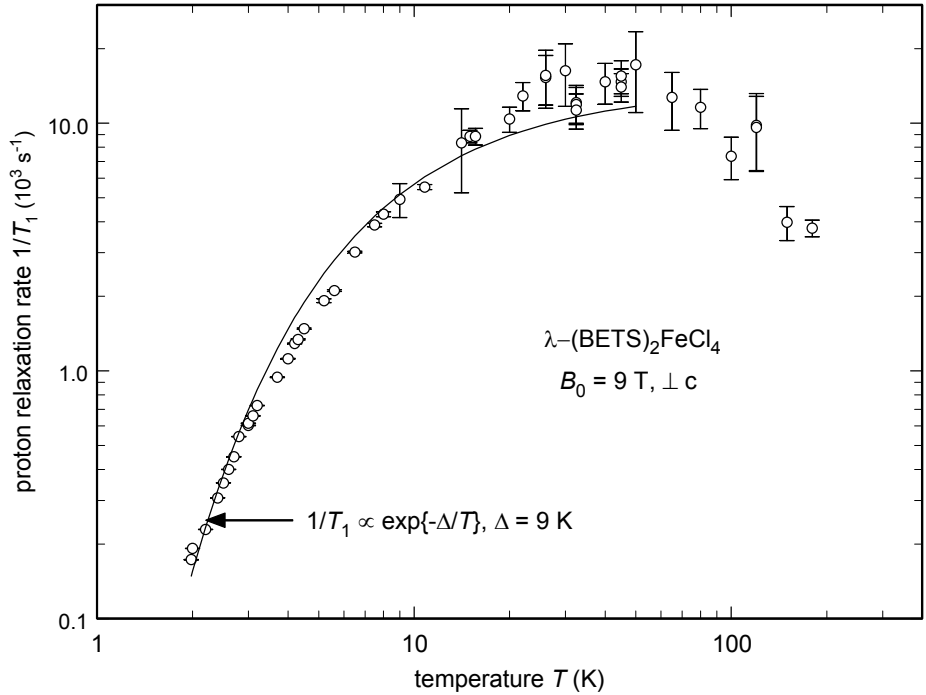

Fig. 3: Proton spin-lattice relaxation rate $1/T_1$ for a single crystal of $\lambda$-$(BETS)_2FeCl_4$ at $B_0 = 9.000$ T in the $a$-$c$ plane perpendicular to the $c$-axis. The solid line indicates a simple energy gap behavior with an energy gap of 9 K.

Several features are evident from the data. From 2 – 20 K, there is a rapid increase over two decades that is within 25 % of the indicated energy-gap behavior. In the range 50 – 180 K, there is a drop by about a factor of 3 in $1/T_1$. There is no clear indication of the PM-AFM transition in the $1/T_1$ measurements. Even though most of the measurements are in the PM phase, there is no evidence of the Korringa-type behavior $T_1T$ = constant that occurs for relaxation by conduction electrons in a metal [9].

It should be pointed out that because of the progressive broadening of the line, even though a serious attempt to measure $1/T_1$ for the same part of the spectrum has been made, some of the $T$-dependence of $1/T_1$ might be caused by including different parts of the spectrum at different values of $T$. This point can be clarified by measurements of $1/T_1$ across the spectrum.

Although a quantitative analysis of the results in Fig. 3 is beyond the scope of this paper, there are several features that can be interpreted qualitatively. One of the most important ones is that all of the values of $1/T_1$ are quite large for the protons. Because the protons are located at the end of the BETS molecules [10] and their atomic number is small, it is expected that the proton hyperfine coupling to the conduction electrons is weak and that the corresponding Korringa relaxation rate should be much smaller than that shown in Fig. 3. We therefore attribute the dominant mechanism for $1/T_1$ of the protons to be the dipole field fluctuations of the $Fe^{3+}$ ions. This conclusion is supported by the spectra shown in Fig 1, where it is seen that the value of the $Fe^{3+}$ dipolar field at the protons is quite large.

An interesting aspect of $1/T_1$ is that as the static magnetization and spectral width become much larger below 30 K (Fig. 1), $1/T_1$ becomes much smaller. This reflects the fact that the proton spectrum measures the local static component of the $Fe^{3+}$ dipolar field parallel to the external field, while $1/T_1$ is determined by its fluctuations perpendicular to the external field [11]. As the $Fe^{3+}$ magnetization becomes saturated at low $T$, is expected that the fluctuations responsible for $1/T_1$ will "freeze out." Two important quantities we expect to show a significant temperature dependence are: (1) The amplitude of the fluctuations and (2) their correlation time. A viable model for $1/T_1$ needs to include both of them and consider fluctuations from both spin-spin and spin-lattice interactions of the $Fe^{3+}$ ions. It is reasonable to expect that in the low temperature limit the corresponding fluctuation rates will have an energy activated behavior that is related to the Zeeman splitting of the $Fe^{3+}$ spins. For a field of 9 T and $g = 2$, this splitting is 12.1 K, which is not far from the value 9 K that comes close to fitting the data. The nearness of these values suggests that saturation of the $Fe^{3+}$ spins in the field at low $T$ plays a significant role in the behavior of the proton $1/T_1$.

SPIN-ECHO DECAY RATE - The proton spin-echo decay rate is shown in Fig. 4, where $1/T_2$ is plotted as a function of $T$ for a single crystal of $\lambda$-(BETS)$_2$FeCl$_4$ at 9.000 T perpendicular to the $c$-axis and the solid line connects the experimental points. These data are for the same part of the spectrum as those shown for $1/T_1$ in Fig. 3. In the region of the PM-AFI transition near 3-4 K at 9 T there is a rapid drop in $1/T_2$ to the value $1.6 \times 10^4$ s$^{-1}$. In the PM phase, the values are substantially larger and have a peak near 50 K. This drop at the transition shows that unlike $1/T_1$, $1/T_2$ is sensitive to the PM-AFI transition.

A reasonable expectation is that the main contributions to $1/T_2$ are the spin-spin interactions between the protons, the fluctuations of the $Fe^{3+}$ dipole field along the applied field on the time scale of $T_2$, and a contribution from $1/T_1$ in the peak near 50 K. Since the proton spin-spin contribution is expected to be temperature independent, we attribute most of the increase in $1/T_2$ above the PM-AFI transition to slow fluctuations of the $Fe^{3+}$ dipole field. There could, of course, be additional contributions associated with the change in the NMR spectrum shown in Fig. 1.

A quantitative interpretation of $1/T_2$ will require a substantial amount of theoretical analysis that is well beyond the scope of this paper.

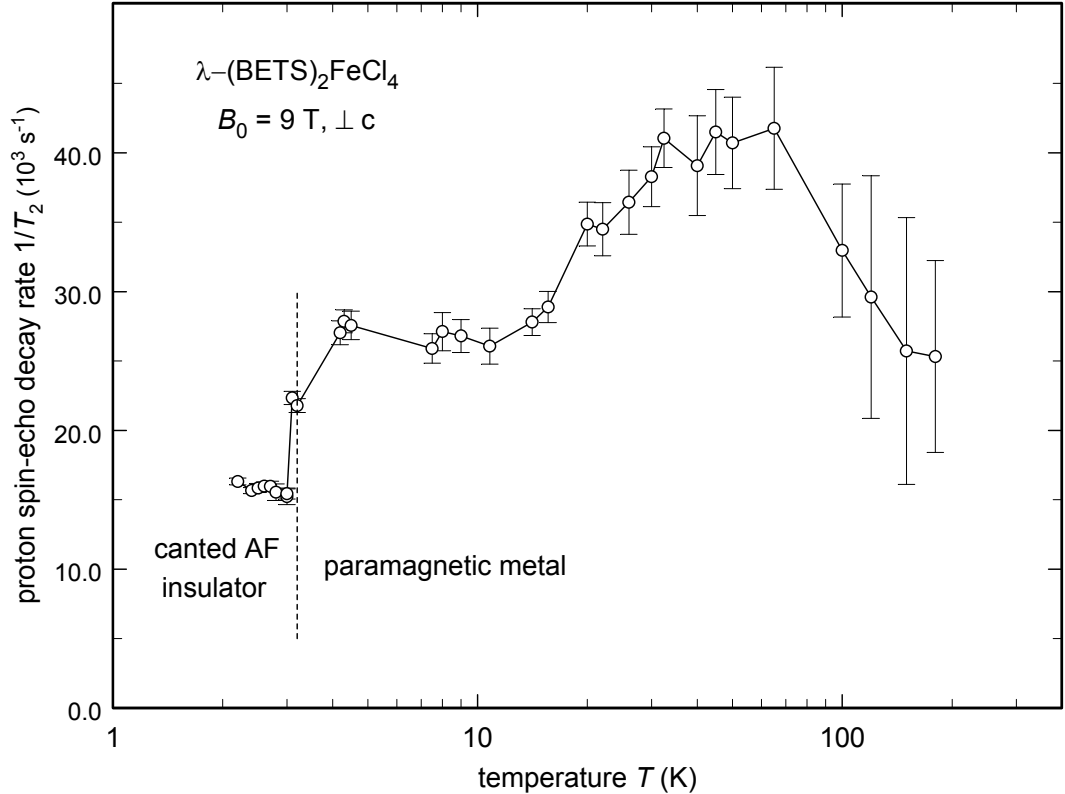

Fig. 4: Proton spin-echo decay rate $1/T_2$ for a single crystal of $\lambda$-(BETS)$_2$FeCl$_4$ at $B_0$ = 9.000 T in the *a-c* plane perpendicular to the *c*-axis. The solid line connects the points and the vertical dashed line shows the approximate position of the PM-AFI transition.

4. CONCLUSIONS

Preliminary NMR measurements of a small single crystal of the two-dimensional organic conductor $\lambda$-(BETS)$_2$FeCl$_4$ at 9 T over the temperature range 1-180 K are reported. They include the paramagnetic metal phase above about 4 K and the antiferromagnetic insulating phase at lower temperature. The steps used to overcome the problem of the small size of the sample ($\sim$3 μg, $\sim$5 × 10$^{16}$ proton spins) and the corresponding interference from spurious proton signals are described. Measurements of the proton spectrum, the proton spin-lattice relaxation rate, and the proton spin-echo decay rate are presented and interpreted qualitatively. Most of these quantities are dominated by the static and fluctuation properties of the Fe3+ spins.

ACKNOWLEDGEMENTS - We thank S.E. Brown, F. Zamborszky, G. Gaidos, and W.Q. Yu for help with this research. The part done at UCLA was supported by NSF Grant DMR-0072524 and that at Indiana by the Petroleum Research Fund ACS-PRF 33912-AC1.

REFERENCES

[1] S. Uji, H. Kobayashi, L. Balicas, and James. S. Brooks, Adv. Mater. **14**, 243 (2002), and references cited therein.
[2] H.N. Bachmann and I.F. Silvera, J. Mag. Res, . **162**, 417 (2003).
[3] S. Endo, T. Goto, T. Fukase, H. Matsui, H. Uozaki, H. Tsuchiya, E. Negishi, Y. Ishizaki, Y. Abe and N. Toyota, J. Phys. Soc. Jpn., **71**, 732 (2002).
[4] . K. Montgomery, T. Burgin, T. Miebach, D. Dunham, and J. C. Huffman, Mol. Cryst. Liq. Cryst. 284, 73 (1996).
[5] .G. Clark, Rev. Sci. Instr. **35**, 316 (1964).
[6] .I. Hoult and R. E. Richards, J. Mag. Res. **24**, 71, (1976).
[7] .G.Clark, M.E. Hanson, F. Lefloch, and P. Ségransan, Rev. Sci. Instr. **66**, 2453-64 (1995).
[8] .W. Ashcroft and D.N. Mermin, "Solid State Physics," Holt, Reinhart, and Winston, New York, 1976, p. 718.
[9] .P. Slichter, "Principles of Magnetic resonance, 3rd edition", Springer-Verlag, Berlin, 1990, Section 5.3.
[10] Hyao Kobayashi, Hideto Tomita, Toshio Naito, Akiko Kobayashi, Fumiko Sakai, Tokuko Watanabe, and Patrick Cassoux, J. Am. Chem. Soc. **118**, 368 (1996).
[11] Reference 9, Ch. 5.